\documentclass[notitlepage,letterpaper,showpacs,preprintnumbers,amsmath,amssymb,nofootinbib,twocolumn]{revtex4-1}

\usepackage{amssymb}
\usepackage{graphicx}
\usepackage{amsmath}
\usepackage{multirow}
\usepackage{verbatim}
\usepackage[usenames,dvipsnames]{color}
\usepackage[bookmarks=true, bookmarksnumbered=true, colorlinks=true,
  urlcolor=darkblue, citecolor=darkgreen, linkcolor=darkred,
  breaklinks=true]{hyperref} 
\usepackage{xspace}

\definecolor{darkblue}{rgb}{0,0,0.5}
\definecolor{darkgreen}{rgb}{0.1,0,0.3}
\definecolor{darkred}{rgb}{0.6,0,0}

\begin{document}

\preprint{IFIC/14-22}


\title{On the flavor composition of the high-energy neutrino events in
  IceCube} 

\author{Olga Mena}
\email{omena@ific.uv.es}
\author{Sergio Palomares-Ruiz}
\email{sergiopr@ific.uv.es}
\author{Aaron C. Vincent}
\email{vincent@ific.uv.es}
\affiliation{Instituto de F\'{\i}sica Corpuscular (IFIC)$,$
 CSIC-Universitat de Val\`encia$,$ \\  
 Apartado de Correos 22085$,$ E-46071 Valencia$,$ Spain}

\begin{abstract}
The IceCube experiment has recently reported the observation of 28
high-energy ($> 30$~TeV) neutrino events, separated into 21 showers
and 7 muon tracks, consistent with an extraterrestrial origin. In this
letter we compute the compatibility of such an observation with possible
combinations of neutrino flavors with relative proportion
($\alpha_e:\alpha_\mu:\alpha_\tau$)$_\oplus$.  Although the 7:21
track-to-shower ratio is naively favored for the canonical
($1:1:1$)$_\oplus$ at Earth, this is not true once the atmospheric
muon and neutrino backgrounds are properly accounted for.  We find
that, for an astrophysical neutrino $E_{\nu}^{-2}$ energy spectrum,
($1:1:1$)$_\oplus$ at Earth is disfavored at 81\%~C.L.  If this
proportion does not change, 6 more years of data would be needed to
exclude ($1:1:1$)$_\oplus$ at Earth at $3\sigma$~C.L.  Indeed, with the
recently-released 3-year data, that flavor composition is excluded at
92\%~C.L.  The best-fit is obtained for ($1:0:0$)$_\oplus$ at Earth,
which cannot be achieved from any flavor ratio at sources with
averaged oscillations during propagation.  If confirmed, this result
would suggest either a misunderstanding of the expected background
events, or a misidentification of tracks as showers, or even more
compellingly, some exotic physics which deviates from the standard
scenario.
\end{abstract}


\maketitle

{\it \textbf{Introduction ---}}
An all-sky search by the IceCube collaboration has recently revealed
the detection of 28 veto-passing events (7 tracks and 21 showers)
between 30~TeV and 1.2~PeV, over a 662~day period, from May 2010 to May
2012~\cite{Aartsen:2013jdh}.  This rate is inconsistent with
atmospheric neutrinos and muons alone, with a significance of
$4.1\sigma$, pointing to a major extraterrestrial component.
Identifying the sources of such a neutrino flux requires dedicated
analyses of the observed events, which include the study of their
energy distribution, their correlation with photons and/or protons,
their arrival direction and their flavor composition.  In this letter
we perform, for the first time, the study of the flavor composition of
the 28 observed events.

The atmospheric neutrino and muon background is expected to be
$10.6^{+5.0}_{-3.6}$ events, of which 8.6 are expected to be
tracks~\cite{Aartsen:2013jdh}.  With only 7 observed tracks, this
implies that the extraterrestrial component overwhelmingly produces
showers inside the detector.  However, this largely departs from the
canonical expectation (see Ref.~\cite{Chen:2013dza}, though).
Astrophysical neutrinos are commonly modeled as the decay products of
pions, kaons and secondary muons produced by (photo)hadronic
interactions. As a result, the expectation for the neutrino flavor
ratio at the source\footnote{We use the subscript ``$\oplus$'' to
  denote the flavor composition as observed by the detector at Earth,
  whereas ``$S$'' represents the composition at the location of the
  astrophysical sources, before any propagation effect takes place.} is
$(\alpha_{e, S}:\alpha_{\mu, S}:\alpha_{\tau, S})=$($1:2:0$)$_S$.
Decoherence occurs after propagating over astronomical distances, 
meaning that oscillations are averaged and this ratio becomes
$(\alpha_{e, \oplus}:\alpha_{\mu, \oplus}:\alpha_{\tau, \oplus}) =$
($1:1:1$)$_\oplus$ at Earth~\cite{Learned:1994wg}.  This is given
explicitly by the measured structure of the neutrino mixing 
matrix~\cite{Tortola:2012te, Fogli:2012ua, GonzalezGarcia:2012sz}, and
leads to a non-negligible component of astrophysically-sourced tracks.
Deviations of the neutrino flavor ratios from this canonical
expectation have been discussed in the literature, as the default
diagnostic of standard effects (including meson energy losses or muon
polarization~\cite{Rachen:1998fd, Kashti:2005qa, Kachelriess:2006fi,
  Lipari:2007su, Pakvasa:2007dc, Hummer:2010ai}), neutron
decays~\cite{Anchordoqui:2003vc}, deviations from tribimaximal
mixing~\cite{Athar:2000yw, Beacom:2003zg, Serpico:2005bs,
  Lipari:2007su, Pakvasa:2007dc, Esmaili:2009dz, Choubey:2009jq,
  Fu:2012zr, Chatterjee:2013tza}, neutrino matter effects in the
source~\cite{Mena:2006eq} and other more exotic
scenarios~\cite{Athar:2000yw, Crocker:2001zs, Beacom:2002vi,
  Barenboim:2003jm, Beacom:2003nh, Beacom:2003eu, Esmaili:2009fk,
  Bhattacharya:2009tx, Bhattacharya:2010xj, Baerwald:2012kc,
  Pakvasa:2012db}.

\begin{figure}[t]
\includegraphics[width=.5\textwidth]{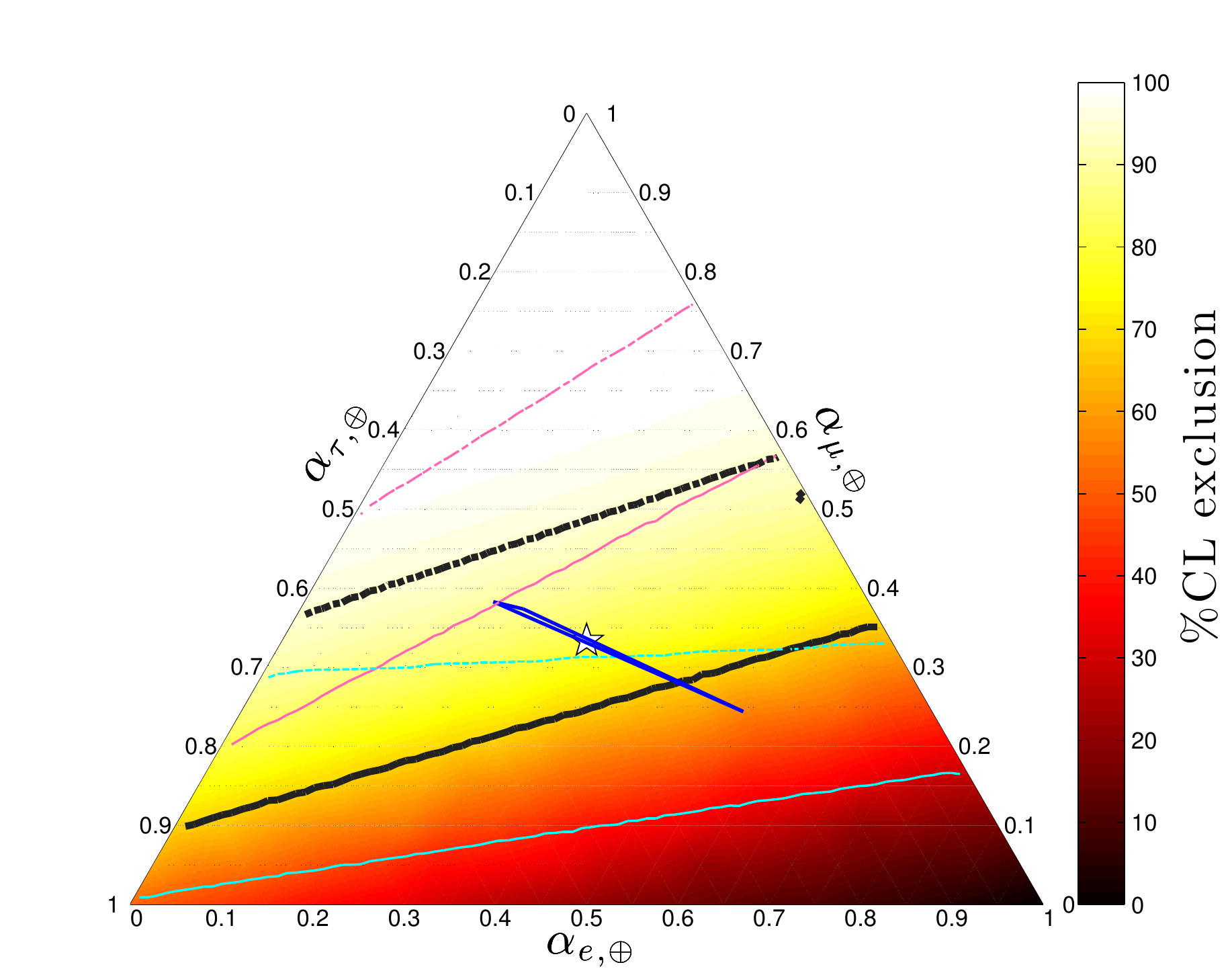}
\caption{Ternary plot of the exclusion C.L. for all possible flavor
  combinations ($\alpha_{e, \oplus}: \alpha_{\mu,\oplus}:
  \alpha_{\tau,\oplus}$) as seen at Earth, given the 7 tracks and 21
  showers observed at IceCube.  The lower right corner corresponds to 
  100\% electron neutrinos, the upper corner is 100\% muon neutrinos,
  and the lower left corner to 100\% tau neutrinos.  The central sliver
  outlined in blue corresponds to the possible flavor combinations for
  astrophysical neutrinos, after oscillations have been averaged
  during propagation.  The best-fit is the darkest point,
  ($1:0:0$)$_\oplus$.  The white star corresponds to
  ($1:1:1$)$_{\oplus}$, which is expected from a ($1:2:0$)$_S$
  combination at the source.  The color scale indicates the exclusion
  C.L. given an $E_{\nu}^{-2}$ spectrum of incoming neutrinos.  Solid
  (dashed) lines show 68\%~C.L. (95\%~C.L.) contours, cyan for
  $E_{\nu}^{-1}$, thick black for $E_{\nu}^{-2}$ and pink for
  $E_{\nu}^{-3}$ spectra.}   
\label{fig:earth}
\end{figure}

\begin{figure}[t]
\includegraphics[width=.5\textwidth]{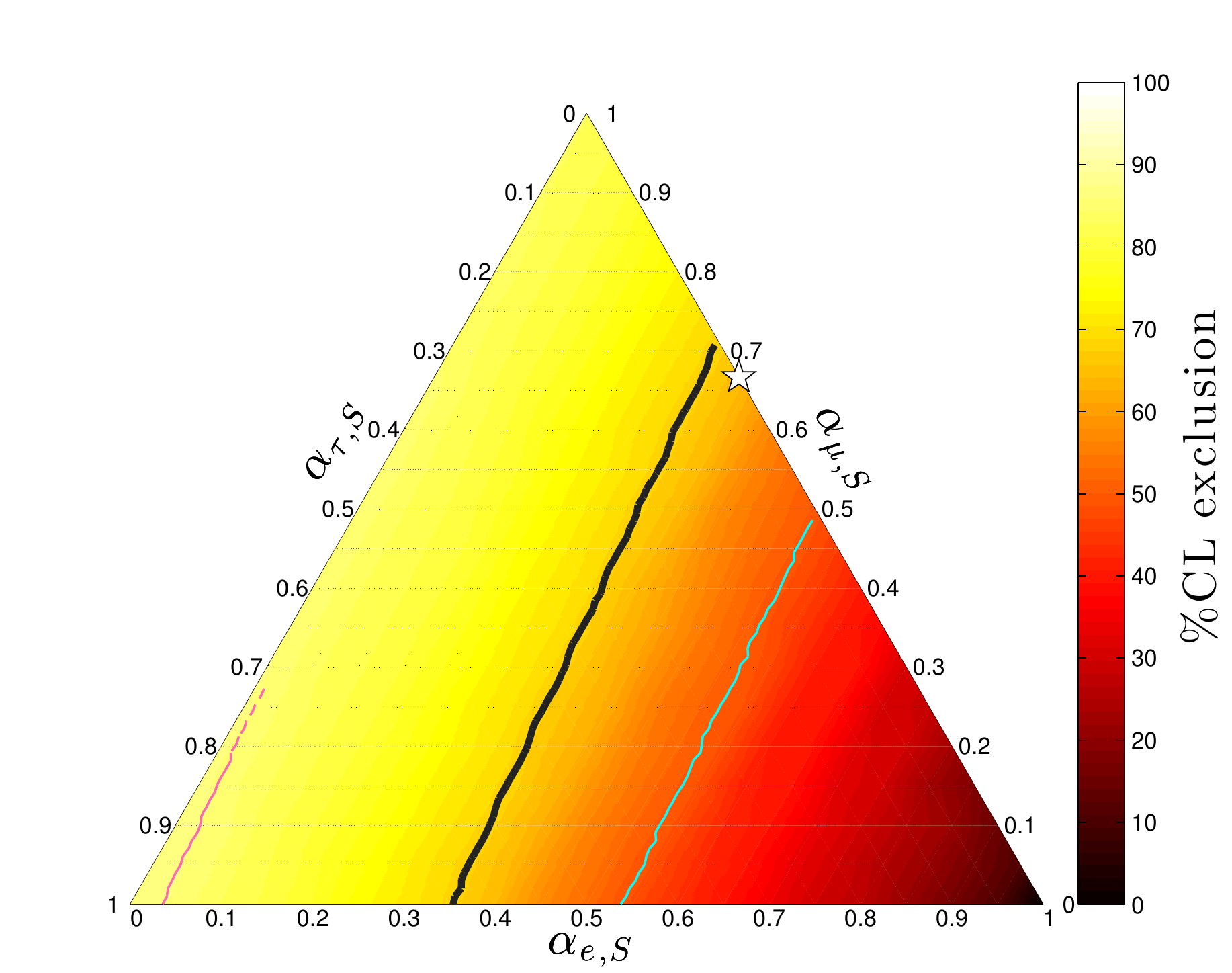}
\caption{Same as Fig.~\ref{fig:earth}, but for $\{\alpha_{i,S}\}$ at
  the source and assuming the signal neutrinos are astrophysical and
  oscillation probabilities are in the averaged regime, i.e., the
  parameter space is restricted to the blue sliver shown in
  Fig.~\ref{fig:earth}.  The best-fit is the darkest point,
  ($1:0:0$)$_S$.  The white star corresponds to the ($1:2:0$)$_S$ flavor
  combination.  Standard flavor compositions lie within a narrow band
  along the right side of the triangle.  Note that all combinations
  are allowed at 95\%~C.L. for the three spectra, and even at
  68\%~C.L. for $E_{\nu}^{-3}$.} 
\label{fig:source}
\end{figure}

Below a few PeV, neutrino flavor ratios can be inferred from two event 
topologies: muon tracks, associated with the \v{C}erenkov light of a
propagating muon, and electromagnetic or hadronic showers.  In this 
letter, we assess the probability of observing the track-to-shower
ratio seen by the IceCube neutrino telescope as a function of the
signal neutrino composition. We consider two parameter spaces: first
the full range ($\alpha_e:\alpha_\mu:\alpha_\tau$)$_\oplus$ at the
detector, and second the restricted range allowed after averaging
oscillations during propagation from astrophysical sources.  We first
outline the calculation of the muon track and shower event rates in
IceCube, after which we describe our statistical approach.  Then, we
present and discuss our results, summarized in Figs.~\ref{fig:earth}
and~\ref{fig:source}.  We show that, after accounting for the expected 
backgrounds, the canonical scheme ($1:1:1$)$_\oplus$ is excluded at the
81\% confidence level (C.L.) for an $E_{\nu}^{-2}$ spectrum.  Finally,
we note that the new 3-year data follow a similar proportion of
tracks and showers~\cite{Aartsen:2014gkd}, which increases the level of
exclusion of the canonical scheme to the 92\%~C.L..

{\it \textbf{Neutrino events in IceCube ---}}
The 28 IceCube events consist of two type of event topologies: muon
tracks and showers.  In both cases, we consider the deposited energy
to be equal to the sum of the energies of all the showers in the event. 

Showers are induced by both $\nu_e$ and $\nu_\tau$ charge current (CC)
interactions, as well as by neutral current (NC) interactions of
neutrinos of all three flavors.  The total number of showers (sh)
produced by NC interactions for any neutrino (and analogously
antineutrino) flavor $i$ reads
\begin{eqnarray}
\label{eq:nc}
N^{\textrm{sh,NC}}_{\nu_i} & = & T \, N_A \, \int^{\infty}_{E_{\textrm{min}}}
dE_\nu \, M^{\textrm{NC}} (E_\nu) \, Att_{\nu_i}(E_\nu) \,
\frac{d\phi_{\nu_i}(E_\nu)}{dE_\nu} \nonumber\\ 
& & \times \int^{y_{\textrm{max}}}_{y_{\textrm{min}}} dy \,
\frac{d\sigma^{\textrm{NC}} (E_\nu,y)}{dy} ~,
\end{eqnarray}
where $E_\nu y = (E_\nu-E'_\nu)$ is the shower energy and $E'_\nu$ is
the energy of the outgoing neutrino, with $y_{\textrm{min}} =
E_{\textrm{min}}/E_\nu$ and $y_{\textrm{max}} = {\textrm{min}} \{ 1, 
E_{\textrm{max}}/E_\nu \}$.  The minimum (maximum) deposited energy in
this analysis is $E_{\textrm{min}} = 30$~TeV ($E_{\textrm{max}} =
2$~PeV).  The differential NC cross section is
${d\sigma^{\textrm{NC}}}/{dy}$, $T=662$~days, $M^{\textrm{NC}}$ is the
energy-dependent effective detector mass for NC interactions,
$N_A=6.022 \times 10^{23} {\rm g}^{-1}$, $Att_{\nu_i}$ is the
attenuation factor due to the absorption and regeneration of $\nu_i$
when traversing the Earth and $d\phi_{\nu_i}/dE_\nu$ is the neutrino
flux. 

Using the same notation, the total number of CC $\nu_e$ (and
analogously $\bar\nu_e$) induced showers reads   
\begin{eqnarray}
\label{eq:cce}
N^{\mathrm{sh, CC}}_{\nu_e} & = & T \, N_A \, \int^{\infty}_{E_{\textrm{min}}}
dE_\nu \, M^{\textrm{CC}}_{\nu_e} (E_\nu) \, Att_{\nu_e}(E_\nu) \,
\frac{d\phi_{\nu_e}(E_\nu)}{dE_\nu} \nonumber\\   
& & \times \int^1_0 dy \, \frac{d\sigma^{\textrm{CC}}_{\nu_e}
  (E_\nu,y)}{dy} \times \Theta\left(E_{\textrm{max}}-E_\nu\right) ~.
\end{eqnarray}

For $\nu_\tau$ (and analogously for $\bar\nu_\tau$), the total number
of shower events induced by CC interactions with an hadronic tau decay
mode is given by~\cite{Dutta:2000jv} 
\begin{eqnarray}
\label{eq:cctau} 
N^{\textrm{sh,CC-had}}_{\nu_\tau} & = & T \, N_A \,
\int^{\infty}_{E_{\textrm{min}}} dE_\nu \,
M^{\textrm{CC}}_{\nu_\tau} (E_\nu) \, Att_{\nu_\tau}(E_\nu) \,
\frac{d\phi_{\nu_\tau}(E_\nu)}{dE_\nu} \nonumber\\ 
& & \times \int^1_0 dy \, \frac{d\sigma^{\textrm{CC}}_{\nu_\tau}
  (E_\nu,y)}{dy} \int^1_0 dz \, \frac{dn(\tau \rightarrow
  {\textrm{had}})}{dz}  \nonumber \\ 
& & \times \Theta\left(E_\nu(y+(1-y)(1-z)) - E_{\textrm{min}}\right)
\nonumber \\
& & \times \Theta\left(E_{\textrm{max}} - E_\nu(y+(1-y)(1-z)\right)) ~, 
\end{eqnarray}
where the total hadronic shower energy is the sum of the hadronic
energy from the broken nucleon, $E_\nu y$, and the hadronic energy
from the decay, $E_\nu (1-y) (1-z)$, where $z= E^\prime_\nu/E_\tau$,
with $E^\prime_\nu$ the energy of the neutrino from the decay.  The
spectrum of the daughter neutrino in hadronic $\tau$ decays is $dn/dz$.

The number of showers produced by the electronic decay of the tau
lepton, $N^{\textrm{sh,CC-em}}$, is written in a similar way, but the
differential distribution is instead the leptonic distribution with 
$z= E_e/E_\tau$ and the $\Theta$ functions in Eq.~(\ref{eq:cctau}) are
replaced by $\Theta \left(E_\nu(y+(1-y)z\right) - E_{\textrm{min}})
\times \Theta \left(E_{\textrm{max}} - 
E_\nu(y+(1-y)z\right))$~\cite{Dutta:2000jv}.  The total number of  
showers produced by $\nu_\tau$ CC interactions (and equivalently by
$\bar\nu_\tau$), $N^{\textrm{sh,CC}}_{\nu_\tau}$, is the sum of the
purely hadronic and hadronic/electromagnetic showers.   

Tracks are induced by muons from $\nu_\mu$ and $\nu_\tau$ CC
interactions.  The energy deposited in the detector comes dominantly  
from the hadronic shower, so the total number of contained-vertex
track-like (tr) events from $\nu_\mu$ (and analogously from
$\bar\nu_\mu$) is
\begin{eqnarray}
\label{eq:tracks}
N^{\textrm{tr}}_{\nu_\mu} & = & T \, N_A \, \int^{\infty}_{E_{\textrm{min}}}
dE_\nu \, M^{\textrm{CC}}_{\nu_\mu} (E_\nu) \, Att_{\nu_\mu}(E_\nu) \, 
\frac{d\phi_{\nu_\mu}(E_\nu)}{dE_\nu} \nonumber\\ 
& & \times \int^{y_{\textrm{max}}}_{y_{\textrm{min}}} dy \,
\frac{d\sigma^{\textrm{CC}}_{\nu_\mu} (E_\nu,y)}{dy} ~.  
\end{eqnarray}

In addition, muon tracks produced by CC $\nu_\tau$ (and $\bar\nu_\tau$)
interactions, $N^{\textrm{tr}}_{\nu_\tau}$, followed by tau decays
($\tau \to \nu_\tau \nu_\mu\mu$), also contribute to the track rate.
To account for these events the branching ratio of tau decays into muons
is included in an equation analogous to Eq.~(\ref{eq:tracks}).    

For the neutrino and antineutrino differential cross sections we use
the \texttt{nusigma} neutrino-nucleon scattering MonteCarlo
code~\cite{Blennow:2007tw}, which uses the CTEQ6 parton distribution
functions~\cite{Pumplin:2002vw, Pumplin:2005rh}.  We use the IceCube
effective masses $M^{\textrm{CC}}_{\nu_i}$ and
$M^{\textrm{NC}}$~\cite{Aartsen:2013jdh}.  The attenuation factors
have been computed for each flavor and for neutrinos and antineutrinos
independently following Refs.~\cite{Naumov:1998sf, Iyer:1999wu,
  Rakshit:2006yi}.  For simplicity and because typically it only
amounts to a small correction~\cite{Dutta:2002zc}, we have not
considered the secondary $\nu_\mu$ flux produced by $\nu_\tau$
interactions~\cite{Beacom:2001xn}.  The attenuation factor in the
above equations is the average attenuation for the whole sky, and thus 
it only depends on the incoming neutrino energy.  We assume the
astrophysical neutrino flux to be given by the same power law,
$E_{\nu}^{-\gamma}$, for the three neutrino and antineutrino flavors.
Although a detailed analysis using all the spectral information will
be described elsewhere, we note that $\gamma \sim 2$ is the value
favored by IceCube data~\cite{Aartsen:2013jdh}.

{\it \textbf{Statistical analysis ---}}
We denote the fractions of electron, muon and tau neutrinos produced in 
astrophysical sources as $\{\alpha_{i, S}\}$.  After propagation,
averaged neutrino oscillations cause the flavor ratio at Earth to be
$\{\alpha_{j,\oplus}\} = \sum_{k, i} |U_{jk}|^2 \, |U_{ik}|^2
\{\alpha_{i, S}\}$, where $U$ is the neutrino mixing matrix for which
we use the latest $\nu$\textit{fit}
results~\cite{GonzalezGarcia:2012sz}.  For $\{\alpha_{i, S}\} =$
($1:2:0$)$_S$, this yields a flavor ratio at Earth of
($1.04:0.99:0.97$)$_\oplus$, very close to the tribimaximal
expectation, ($1:1:1$)$_\oplus$.   

The total number of events produced by astrophysical neutrinos, for a
given combination $\{\alpha_{i,\oplus}\}$, is 
\begin{eqnarray}
\label{eq:Na}
N_{\rm a}(\{\alpha_{i,\oplus}\}) & = & \alpha_{e, \oplus} \, (
N_{\nu_e}^{\textrm{sh,CC}} + N_{\nu_e}^{\textrm{sh,NC}} ) \nonumber \\  
 &  & + \alpha_{\mu, \oplus} \, (N^{\textrm{tr}}_{\nu_\mu} +
N^{\textrm{sh,NC}}_{\nu_\mu}) \nonumber \\  
 & & + \alpha_{\tau, \oplus} \, (N^{\textrm{tr}}_{\nu_\tau} +
N^{\textrm{sh,CC}}_{\nu_\tau} + N^{\textrm{sh,NC}}_{\nu_\tau}) ~,
\end{eqnarray}
where we implicitly assume the sum of neutrino and antineutrino
events.  The proportion of these events which is expected to produce
muon tracks is 
\begin{equation}
p_{\textrm{a}}^{\textrm{tr}}(\{\alpha_{i, \oplus}\}) = \frac{1}{N_{\rm
    a}(\{\alpha_{i,\oplus}\})} \left( \alpha_{\mu,\oplus} \,
N^{\textrm{tr}}_{\nu_\mu} + \alpha_{\tau,\oplus} \,
N^{\textrm{tr}}_{\nu_\tau} \right) ~,  
\label{ptdef}
\end{equation}
and conversely for showers, $p_{\textrm{a}}^{\rm sh}
(\{\alpha_{i,\oplus}\}) \equiv 1 - p_{\textrm{a}}^{\rm tr}
(\{\alpha_{i,\oplus}\})$.

For the background we consider $b_\mu = 6$ atmospheric muons and
$b_\nu = 4.6$ atmospheric neutrinos~\cite{Aartsen:2013jdh}.  We take
the background events to be Poisson-distributed and only consider
statistical errors.  We note that the lower systematic error quoted by
the IceCube collaboration on the total number of expected background
events is $3.6$, which is comparable to the statistical error for
$10.6$ events.  This could reduce the significance of the exclusion
limits we present below, whereas the upper value of the systematic
error would pull the analysis towards a worse fit for
($1:1:1$)$_\oplus$, thus not affecting our results significantly.    
Additionally, neutrinos from atmospheric charmed meson decays could,
in the benchmark model, represent 1.5 extra background events.  Given
the uncertainty in this prediction (see, e.g.,
Ref.~\cite{Enberg:2008te}), we consider this case separately.  For the
fraction of background showers and tracks in the
$30~\textrm{TeV}-2$~PeV energy range, we use the numbers quoted by the
IceCube collaboration: tracks account for $69\%$ of the conventional
atmospheric neutrino event rate, $19\%$ of the prompt atmospheric
neutrino event rate and $90\%$ of the events induced by atmospheric
muons~\cite{Aartsen:2014gkd}.  We have also checked that the
uncertainties in the ratio of tracks to showers from atmospheric
neutrinos, as computed with different initial fluxes, do not change
our results in a significant way.  For instance, using the high-energy
atmospheric neutrino fluxes of Refs.~\cite{Sinegovsky:2011ab,
  Petrova:2012qf, Sinegovskaya:2013wgm}, the fraction of tracks
induced by the conventional flux is $\sim 50\%$.  This would only
weaken our conclusions by decreasing the exclusion C.L. by a few
percent.

The likelihood of observing $N_{\textrm{tr}}$ tracks and $N_{\textrm{sh}}$
showers, for a given combination $\{\alpha_{i,\oplus}\}$ and a total
number of astrophysical neutrinos $N_{\textrm{a}}$, is 
\begin{align}
\nonumber  
\mathcal{L}(& \{\alpha_{i,\oplus}\}, N_{\textrm{a}}| N_{\textrm{tr}},
N_{\textrm{sh}})  = \\ 
  & e^{- (p_{\textrm{a}}^{\textrm{tr}} N_{\textrm{a}} +
  p_{\mu}^{\textrm{tr}} b_\mu + p_{\nu}^{\textrm{tr}} b_\nu)} \, 
\frac{(p_{\textrm{a}}^{\textrm{tr}} \, N_{\textrm{a}} +
  p_{\mu}^{\textrm{tr}} \, b_\mu + p_{\nu}^{\textrm{tr}} \,
  b_\nu)^{N_{\textrm{tr}}}}{N_{\textrm{tr}}  !} \nonumber \\    
\times & \, e^{- (p_{\textrm{a}}^{\textrm{sh}} N_{\textrm{a}} +
  p_{\mu}^{\textrm{sh}} b_\mu + p_{\nu}^{\textrm{sh}} b_\nu)} \,
\frac{( p_{\textrm{a}}^{\textrm{sh}} \, N_{\textrm{a}} +
  p_{\mu}^{\textrm{sh}} \, b_\mu + p_{\nu}^{\textrm{sh}} \,
  b_\nu)^{N_{\textrm{sh}}}}{N_{\textrm{sh}} !} ~,      
\end{align}
where $p_{\nu}^{\textrm{tr}} = 0.69$ ($p_{\nu}^{\textrm{sh}} = 1 -
p_{\nu}^{\textrm{tr}}$) is the fraction of tracks (showers) in the
atmospheric neutrino background and $p_{\mu}^{\textrm{tr}} = 0.9$
($p_{\mu}^{\textrm{sh}} = 1 - p_{\mu}^{\textrm{tr}}$) is the fraction
of tracks (showers) in the atmospheric muon
background~\cite{Aartsen:2014gkd}.  Since the total number of events  
produced by astrophysical neutrinos is not of interest in this
analysis, $N_{\textrm{a}}$ can be treated as a nuisance parameter and
can be set to the value $N_{\textrm{a}}^{\rm  max} (\{\alpha_{i,\oplus}\})$
which maximizes $\mathcal{L}(\{\alpha_{i,\oplus}\}, N_{\textrm{a}} |
N_{\textrm{tr}}, N_{\textrm{sh}})$ for $\{\alpha_{i,\oplus}\}$, yielding 
$\mathcal{L}_{\rm p}(\{\alpha_{i,\oplus}\} | N_{\textrm{tr}},
N_{\textrm{sh}}) \equiv \mathcal{L}(\{\alpha_{i,\oplus}\},
N_{\textrm{a}}^{\rm max} (\{\alpha_{i,\oplus}\}) | N_{\textrm{tr}},
N_{\textrm{sh}})$.

We construct the log-likelihood ratio
\begin{equation}
\lambda(N_{\textrm{tr}},N_{\textrm{sh}} | \{\alpha_{i,\oplus}\}) = -2 \ln
\left( \frac{\mathcal{L}_{\rm p}(\{\alpha_{i,\oplus}\} | N_{\textrm{tr}},
  N_{\textrm{sh}})}{\mathcal{L}_{\rm p}(\{\alpha_{i,\oplus}\}_{\rm max} |
  N_{\textrm{tr}},N_{\textrm{sh}})} \right) ~,  
\end{equation}
where $\{\alpha_{i,\oplus}\}_{\rm max} $ is the combination of
neutrino flavors that maximizes the likelihood of observing
$N_{\textrm{tr}}$ tracks and $N_{\textrm{sh}}$ showers.  The p-value
for a given combination $\{\alpha_{i,\oplus}\}$ is 
\begin{equation}
p(\{\alpha_{i,\oplus}\}) = \sum_{N_{\textrm{tr}},N_{\textrm{sh}}}
  P(N_{\textrm{tr}}, N_{\textrm{sh}} | \{\alpha_{i,\oplus}\}) ~, 
\label{pvalue}
\end{equation}
where $P(N_{\textrm{tr}}, N_{\textrm{sh}} | \{\alpha_{i,\oplus}\}) \equiv
\mathcal{L}_{\rm p}(\{\alpha_{i,\oplus}\} | N_{\textrm{tr}},
N_{\textrm{sh}})$ is the probability of observing $N_{\textrm{tr}}$
tracks and $N_{\textrm{sh}}$ showers given the flavor ratio
$\{\alpha_{i,\oplus}\}$ and $N_{\textrm{a}}^{\rm max}
(\{\alpha_{i,\oplus}\})$, and the sum runs over all combinations of
$N_{\textrm{tr}}$ and $N_{\textrm{sh}}$ which satisfy
$\lambda(N_{\textrm{tr}}, N_{\textrm{sh}} | \{\alpha_{i,\oplus}\}) > 
\lambda(N_{\textrm{tr}} = 7, N_{\textrm{sh}} =  21 |
\{\alpha_{i,\oplus}\})$.  The test statistic $\lambda$ asymptotically
approaches a $\chi^2$ distribution with two degrees of freedom.  The
p-value can easily be translated into an exclusion C.L.: $C.L.
(\{\alpha_{i,\oplus}\}) = 1 - p (\{\alpha_{i,\oplus}\})$.

{\it \textbf{Results ---}}
Using Eq.~(\ref{pvalue}), we compute the exclusion limits for all
combinations of $\{\alpha_{i,\oplus}\}$, without any restrictions on
the flavor ratios at Earth.  We show the results of Eq.~(\ref{pvalue})
in Fig.~\ref{fig:earth} and provide several exclusions limits for
($1:1:1$)$_\oplus$ at Earth in Tab.~\ref{tab:cltabEarth}.  The color
scale shows the exclusion C.L. assuming an $E_{\nu}^{-2}$ astrophysical
spectrum for all three flavors, which describes well data in the
$30~\textrm{TeV}-2$~PeV energy range~\cite{Aartsen:2013jdh}.  Lines
show the 68\% and 95\%~C.L. limits, which we illustrate for three
different spectra.  The ($1:1:1$)$_\oplus$ scenario is excluded at
81\%~C.L. for an $E_{\nu}^{-2}$ spectrum.  Harder spectra are more
constrained, since a larger flux of $\nu_\mu$'s and $\nu_\tau$'s at 
high energies necessarily leads to the production of more muons.  We 
note that the best-fit point is ($1:0:0$)$_\oplus$, which cannot be
obtained from any flavor ratio at sources assuming averaged
oscillations during propagation.

We now turn to the following question: what happens if we impose the
restriction that the observed non-atmospheric neutrinos are  
extraterrestrial, such that oscillations are averaged during
propagation? In this case, they must be contained within the blue
sliver of Fig.~\ref{fig:earth}, and the event topology data become
less constraining, at the expense of an overall worse fit.  This is
shown in Fig.~\ref{fig:source}, where one can see that ($1:2:0$)$_S$ for
the $E_{\nu}^{-2}$ spectrum is disfavored at 65\%~C.L. with respect to
the best-fit, ($1:0:0$)$_S$, which could be explained, for instance, by
neutron decay sources~\cite{Anchordoqui:2003vc}.  However, we note
that a large fraction of Fig.~\ref{fig:source} is disfavored at
1$\sigma$~C.L. or more with respect to the best-fit in
Fig.~\ref{fig:earth}.  Different exclusion limits for this case are
also presented in Tab.~\ref{tab:cltabEarth}.
  
Beyond the conventional $\pi/K$ atmospheric neutrino background, the 
effect of an atmospheric charm component is shown in
Tab.~\ref{tab:cltabEarth}, where we see that the changes are not
significant.

\begin{table}[t]
\begin{tabular*}{1\linewidth}{@{\extracolsep{\fill}}c | c c c} 
$d \phi_\nu /d E_\nu \propto$ & $E_\nu ^{-1}$ & $E_\nu ^{-2}$ & $E_\nu
  ^{-3} $  \\ \hline 
$\pi/K$            & 96\% (78\%) & 81\% (65\%) & 52\% (36\%) \\
$\pi/K$ + charm    & 95\% (76\%) & 80\% (63\%) & 53\% (37\%) \\ \hline
$\pi/K$ (3-yr data)& 99\% (87\%) & 92\% (77\%) & 70\% (52\%)
\end{tabular*}
\caption{Exclusion limits for the ($1:1:1$)$_{\oplus}$ flavor ratio
  observed at Earth (for the ($1:2:0$)$_S$ flavor ratio at the source
  and assuming averaged oscillations).  The three columns represent
  three possible assumptions for the spectrum of the astrophysical
  neutrinos as a function of their energy.  ``$\pi/K$'' includes the
  conventional atmospheric muon and neutrino background and ``$\pi/K$
  + charm'' additionally includes the benchmark flux of ``prompt''
  neutrinos from the decay of charmed mesons in the atmosphere.  The
  two upper rows refer to the 2-year data~\cite{Aartsen:2013jdh} and
  the last one to the recently released 3-year
  data~\cite{Aartsen:2014gkd}.}
\label{tab:cltabEarth}
\end{table}

{\it \textbf{Discussion ---}} 
Although the statistical power of the high-energy events seen at
IceCube remains low, the 8.6 tracks expected from the atmospheric muon
and neutrino backgrounds allow us to place moderate constraints on the
flavor ratios of the non-background neutrinos.  If these are assumed
to have an $E_{\nu}^{-2}$ energy spectrum and allowed to take any
combination, the ($1:1:1$)$_\oplus$ ratio at Earth is excluded at
81\%~C.L.  If they are constrained to be astrophysically-sourced
and oscillation probabilities are averaged during the propagation to
Earth, this exclusion is reduced to 65\%~C.L.  This is simply due to
the reduction of the parameter space, so the likelihood varies by 
smaller amounts with respect to the full $\{\alpha_{i,\oplus}\}$ space,
leading to a smaller constraining power for the same sample size.  

It is compelling to note that significant limits are potentially at
hand.  Indeed, the new 3-year IceCube data~\cite{Aartsen:2014gkd}
indicate the detection of 9 extra events, of which only 2 are tracks.
Hence, the proportion of tracks and showers after 3 years is similar
to that in the 2-year data.  With an expected background of $8.4 \pm
4.2$ atmospheric muons and $6.6^{+5.9}_{-1.6}$ atmospheric neutrinos,
this implies that, for an $E_{\nu}^{-2}$ spectrum, ($1:1:1$)$_\oplus$ at
Earth [($1:2:0$)$_S$ at source] is excluded at 92\%~C.L. (77\%~C.L.).
For other spectra, 3-year exclusion limits are presented in
Tab.~\ref{tab:cltabEarth}.  With the new data, the best-fit at source,
($1:0:0$)$_S$, is excluded with respect to the best-fit at Earth,
($1:0:0$)$_\oplus$, at 75\%~C.L. for an $E_{\nu}^{-2}$ spectrum.  Let us
also note that for the best-fit spectrum quoted by IceCube,
$E_{\nu}^{-2.3}$~\cite{Aartsen:2014gkd}, ($1:1:1$)$_\oplus$ at Earth
[($1:2:0$)$_S$ at source] is excluded at 86\%~C.L. (70\%~C.L.). If the 
ratio of 1 track per 3 showers holds for future observations,
($1:1:1$)$_\oplus$ could be excluded at $3\sigma$~C.L. for an
$E_{\nu}^{-2}$ spectrum after a total of 8 years.  If this trend
continues, we are faced with several potential implications: (a) the
main mechanism of astrophysical neutrino production is \textit{not}
purely hadronic interactions and indeed the best-fit at source is
($1:0:0$)$_S$ indicating an origin in neutron, rather than meson,
decay; (b) no flavor combination at the source provides a good fit to
the data and hence, the observed flavor ratios are due to some
non-standard effect which favors a dominant $\nu_e$ composition at
Earth, for instance as in some scenarios of neutrino decay, {\sl CPT}
violation or pseudo-Dirac neutrinos~\cite{Crocker:2001zs,
  Beacom:2002vi, Barenboim:2003jm, Baerwald:2012kc, Pakvasa:2012db};
(c) the atmospheric background has been overestimated; or (d) some
tracks have been misidentified as showers.   

The 28 IceCube events have opened the door to the era of neutrino
astronomy.  Even with such a small sample, the event topology provides
compelling information on the production, propagation and detection of
neutrinos at high energies.  Future data has the potential to firmly
establish the origin and composition of these neutrinos.

{\it \textbf{Acknowledgments ---}} 
We thank Claudio Kopper and Nathan Whitehorn for clarifying details
about the IceCube data and for providing us with the effective masses,
and Sergei Sinegovsky for providing us with tabulated high-energy
atmospheric neutrino fluxes.  We also thank Glen Cowan and Miguel
Nebot for enlightning discussions on statistics and Pilar Hern\'andez,
Carlos Pe\~na-Garay, Tom Weiler and Walter Winter for useful comments.
OM is supported by the Consolider Ingenio project CSD2007--00060, by
PROMETEO/2009/116, by the Spanish Grant FPA2011--29678 of the MINECO.
SPR is supported by a Ram\'on y Cajal contract and by the Spanish
MINECO under grant FPA2011-23596.  ACV is supported by FQRNT and
European contract FP7-PEOPLE-2011-ITN.  The authors are also partially
supported by PITN-GA-2011-289442-INVISIBLES.  SPR is also partially
supported by the Portuguese FCT through the projects
CERN/FP/123580/2011, PTDC/FIS-NUC/0548/2012 and CFTP-FCT Unit 777
(PEst-OE/FIS/UI0777/2013), which are partially funded through POCTI
(FEDER).

\bibliographystyle{apsrev4-1}
\bibliography{flavors}

\end{document}